\documentclass[
    aps, prb,
    reprint,
    superscriptaddress,
    showkeys,
    amsmath,
    amssymb,
    a4paper, 8pt
]{revtex4-2}

\usepackage{geometry}
\usepackage{graphicx}
\usepackage{siunitx}
\DeclareSIUnit\angstrom{\text{Å}}

\usepackage{xcolor}
\definecolor{petroff1}{HTML}{3f90da}
\definecolor{petroff2}{HTML}{ffa90e}
\definecolor{petroff3}{HTML}{bd1f01}
\definecolor{petroff4}{HTML}{94a4a2}
\definecolor{petroff5}{HTML}{832db6}
\definecolor{petroff6}{HTML}{a96b59}
\definecolor{petroff7}{HTML}{e76300}
\definecolor{petroff8}{HTML}{b9ac70}
\definecolor{petroff9}{HTML}{717581}
\definecolor{petroff10}{HTML}{92dadd}

\newcommand{\scatter}[1]{\textcolor{#1}{$\bullet$}}

\begin{document}

\title{\textbf{Vapor--solid--solid growth of single-walled carbon nanotubes}}

\author{Daniel Hedman}
\email{daniel@hedman.science}
\affiliation{Center for Multidimensional Carbon Materials (CMCM), Institute for Basic Science (IBS), Ulsan, 44919, Republic of Korea}

\date{December 30, 2025}

\begin{abstract}
    Single-walled carbon nanotubes (SWCNTs) are promising for nanoscale electronics and photonics, but practical deployment requires chirality control. Most catalytic chemical vapor deposition (CCVD) growth of SWCNTs proceeds on liquid metal nanoparticles via a vapor--liquid--solid (VLS) mechanism and yields broad chirality distributions, whereas improved selectivity has been reported for high-melting-point crystalline catalysts, suggesting vapor--solid--solid (VSS) growth. However, the atomistic mechanism and kinetic of VSS SWCNT growth remain unclear. Here it is shown, using machine-learning interatomic potential-driven molecular dynamics on rhenium nanoparticles, that VSS growth is diffusion-limited and governed by facet-dependent surface carbon transport coupled to carbon-driven facet reconfiguration without catalyst melting. Surface diffusion is up to $\sim 50\times$ slower than carbon diffusion in liquid iron, imposing a strict upper bound on sustainable carbon supply and producing a narrow growth window: insufficient transport drives carbon accumulation and multiple nucleation, whereas higher temperatures favor graphitic encapsulation. In contrast to defect-free VLS growth, defects persist, indicating slow defect healing, and equilibrium simulations reveal suppressed edge configurational entropy with stabilization of zigzag-rich, Klein-decorated edges. Together, these results establish facet evolution and surface diffusion as joint regulators of diffusion-limited VSS growth and motivate stringent control of temperature and carbon supply.
\end{abstract}


\maketitle

\section{Introduction}\label{sec:introduction}
 
Carbon nanotubes (CNTs) are hollow cylinders of $sp^2$-bonded carbon atoms arranged in a hexagonal lattice, with a single wall or multiple nested walls. For single-walled CNTs (SWCNTs), electronic and optical properties are set by chirality, described by the chiral indices $(n, m)$ that define the orientation of the lattice relative to the tube axis~\cite{dresselhaus1995physics}. Chirality is set during the initial stages of synthesis (growth) as the SWCNT-cap is formed, but subsequent incorporation of defects (non-hexagonal rings, e.g., pentagons and heptagons) can change the effective chirality.

Catalytic chemical vapor deposition (CCVD) remains the dominant synthesis route for SWCNTs and commonly employs transition-metal nanoparticles (NPs) as catalysts. For widely used catalysts such as iron~\cite{he2012diameter} and nickel~\cite{hofmann2007situ}, as well as alloys such as cobalt--molybdenum~\cite{kitiyanan2000controlled}, iron--nickel~\cite{chiang2009linking}, and iron--cobalt~\cite{hu2026structuresironcobaltbimetallic}, nanoscale melting-point depression combined with elevated CCVD temperatures (typically above \SI{800}{\kelvin}) results in liquid catalysts. Under these conditions, SWCNT growth is commonly described by a vapor--liquid--solid (VLS) mechanism~\cite{saito1994singlewall} and typically results in broad chirality distributions~\cite{chiang2009linking,he2012diameter,otsuka2022universal,hedman2024dynamics,sun2025chiralitydependent}.

Improved chirality or diameter selectivity has been reported for growth on high-melting-point catalysts that remain solid during CCVD, including refractory metals such as molybdenum, tungsten, and rhenium~\cite{zhang2020highprecision}; carbide NPs such as molybdenum carbide~\cite{zhang2015diameterspecific} and tungsten carbide~\cite{zhang2017arrays}; and refractory alloy NPs such as cobalt--tungsten~\cite{yang2014chiralityspecific,yang2015growing,yang2016waterassisted} and cobalt--rhenium~\cite{li2022kineticscontrolled}. On such solid-state catalysts, SWCNT growth has been proposed to occur via a vapor--solid--solid (VSS) mechanism~\cite{yoshida2008atomicscale} rather than VLS. Despite growing experimental evidence for VSS growth, the underlying atomistic growth mechanisms remain poorly understood.

In contrast, machine-learning interatomic potential (MLIP)-driven molecular dynamics (MD) has recently provided atomistic insight into VLS growth of SWCNTs on iron~\cite{hedman2024dynamics,kohata2025edge} and cobalt~\cite{sun2025chiralitydependent} catalysts. These growth simulations have revealed a highly dynamic tube--catalyst interface with large fluctuations in SWCNT-edge structures, i.e., high edge configurational entropy. Interface defects were found to form stochastically but can, under appropriate conditions, be effectively healed before becoming trapped in the tube wall, enabling defect-free growth. Simulations of VLS growth on cobalt have indicated that continued growth after incorporation of defects can act as a diameter-control pathway, narrowing the chirality distribution. However, this type of mechanistic understanding remains limited for VSS growth of SWCNTs, partly due to the lack of growth simulations on solid-state catalysts at relevant length and timescales. The present work takes a first step toward addressing this gap by using MLIP-driven MD simulations to study SWCNT growth on solid-state rhenium NPs.

Specifically, catalyst stability under growth conditions is investigated, facet-dependent carbon transport is quantified, and \si{\micro\second}-scale MD growth simulations are used to link transport and interface defect kinetics to cap nucleation and tube elongation. In addition, equilibrium annealing of SWCNTs attached to two prominent facets is used to probe how the solid-state catalyst affects the edge thermodynamics, suppressing the configurational entropy of the SWCNT-edge. Overall, VSS SWCNT growth on rhenium is governed by carbon-driven facet reconfiguration together with facet-dependent surface diffusion, leading to a diffusion-limited growth regime and markedly slower interface and defect kinetics compared to VLS. These findings provide an atomic-scale mechanistic basis for understanding SWCNT growth on solid-state catalysts and motivate tight experimental control of temperature and carbon supply under diffusion-limited VSS conditions.

\section{Results and discussion}\label{sec:results}

Machine-learning interatomic potentials enable atomistic simulations at length and timescales approaching experiment by learning energies, forces, and virials from first-principles reference data, typically density functional theory (DFT) calculations. After training, MLIPs can be used to drive molecular dynamics simulations with computational cost comparable to empirical force fields while retaining near-DFT accuracy. A central challenge in MLIP construction is to obtain a dataset that is both high-quality and sufficiently diverse.

\begin{figure}
  \centering
  \includegraphics[width=\columnwidth]{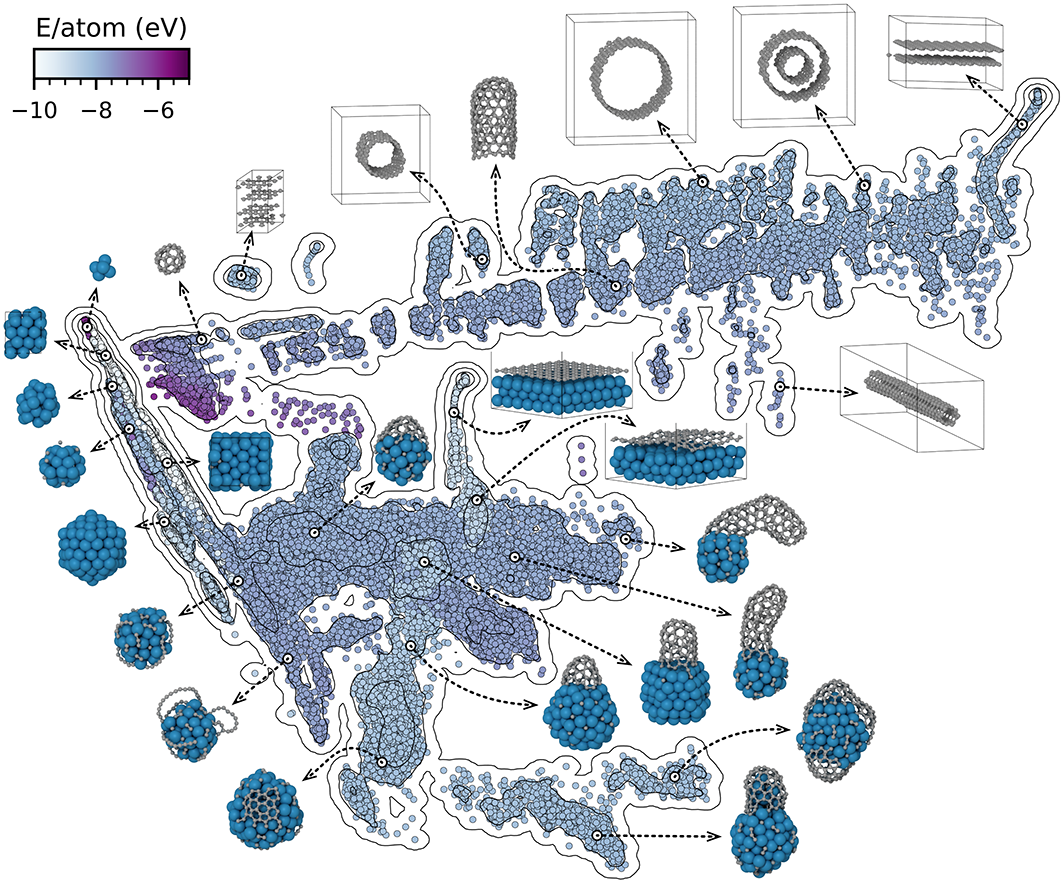}
  \caption{\label{fig:sketch-map} Sketch-map representation of the Re--C dataset where each colored point represents an individual atomic configuration (structure) and contour lines show structure density. Total numbers of structures are 86 474 (training) and 9 549 (validation). Each point is positioned using principal component analysis of the learned descriptors for the structures and colored by the corresponding energy. Example structures from different regions on the sketch-map are shown to provide insight into the diversity of the dataset. Blue atoms depict rhenium and gray atoms carbon.}
\end{figure}

Here, a Re--C dataset was generated by active learning~\cite{podryabinkin2017active,smith2018sampling} and is summarized by the sketch-map representation~\cite{ceriotti2011simplifying} in Fig.~\ref{fig:sketch-map}. Each point corresponds to a unique atomic configuration, positioned by principal component analysis of the learned descriptors of the local atomic environments. The observed clustering indicates that structurally similar configurations map to nearby regions while dissimilar configurations separate, consistent with a broad coverage of configurations relevant to SWCNT growth on rhenium.

A neuroevolution potential (NEP)~\cite{song2024generalpurpose} was trained on the Re--C dataset using a 90\%/10\% split into training and validation sets. NEP\textsubscript{Re--C} is highly accurate with energy and force root-mean-square errors of \SI{11.44}{\milli\electronvolt} per atom and \SI{398.7}{\milli\electronvolt\per\angstrom}, respectively, see Fig. S1, comparable to previous MLIPs used for SWCNT growth~\cite{hedman2024dynamics,sun2025chiralitydependent,kohata2025edge}. Additional verification of NEP\textsubscript{Re--C} is provided in Sec. 1 of the Supporting Information (SI). The computational efficiency of NEP\textsubscript{Re--C} enables up to \SI{150}{\nano\second} per day of simulated SWCNT growth, depending on system size, using the Graphics Processing Units Molecular Dynamics (GPUMD)~\cite{fan2022gpumd,xu2025gpumd} software on an NVIDIA T4 GPU.

\subsection{Nanoparticle stability, facets, and carbon transport}

\begin{figure*}
  \centering
  \includegraphics[width=\textwidth]{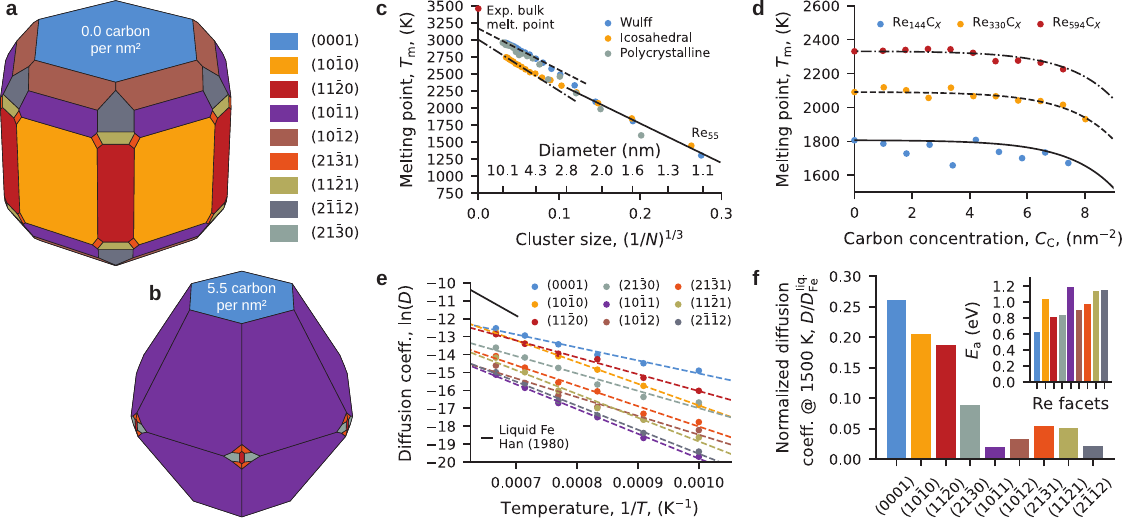}
  \caption{\label{fig:cluster} The melting point, $T_\text{m}$, and facet-dependent carbon diffusion on rhenium nanoparticles (NPs). Wulff constructs of rhenium NPs, \textbf{a} without and \textbf{b} with surface carbon. \textbf{c} $T_\text{m}$ as a function of NP size, $N$, for \scatter{petroff1} Wulff, \scatter{petroff2} icosahedron and \scatter{petroff4} polycrystalline NPs; \scatter{petroff3} shows the experimental bulk melting point of rhenium. Here the solid, dash-dot and dashed lines are linear regressions fitted to \scatter{petroff1} and \scatter{petroff2} for $N < 455$ atoms, to \scatter{petroff2} for $N > 2915$ and to \scatter{petroff1} for $N > 2915$, respectively. \textbf{d} $T_\text{m}$ as a function of surface carbon concentration, $c$, for three different-sized Wulff NPs \scatter{petroff1} Re\textsubscript{144}C\textsubscript{X}, \scatter{petroff2} Re\textsubscript{330}C\textsubscript{X} and \scatter{petroff3} Re\textsubscript{594}C\textsubscript{X}. Here the solid, dash-dot and dashed lines are trend lines, $T_\text{m}^{c=0} - \left(e^{a c} - 1\right)$, fitted to each NP size. \textbf{e} Temperature-dependent carbon diffusion coefficient, $D$, on rhenium facets; here the black line shows the carbon diffusion coefficient in liquid iron~\cite{han1980determination}. \textbf{f} Carbon diffusion coefficient on rhenium facets at \SI{1500}{\kelvin} normalized by the carbon diffusion coefficient in liquid iron at the same temperature. Here the inset shows the facet-dependent activation energies, $E_a$, for carbon diffusion obtained from \textbf{e}. In \textbf{e} and \textbf{f} the carbon concentration is $c = \SI{5.5}{\nano\meter^{-2}}$.}
\end{figure*}

For vapor--solid--solid growth, the catalyst is expected to remain crystalline. Because SWCNT growth typically needs very small NPs---to avoid multi-walled CNTs and carbon nanofibers---the melting point, $T_\text{m}$, of catalysts must be high even at nanoscale sizes.

Here, melting simulations of polycrystalline, icosahedral, and Wulff rhenium NPs were performed using NEP\textsubscript{Re--C}. As shown in Fig.~\ref{fig:cluster}c, even very small $N = 55$ atoms ($\sim\SI{1}{\nano\meter}$) rhenium NPs exhibit high melting points above \SI{1000}{\kelvin}. For NP sizes closer to those used by Zhang \emph{et al.}~\cite{zhang2020highprecision} to experimentally grow SWCNTs on rhenium, $\sim\SI{1.8}{\nano\meter}$ diameter, the melting point exceeds \SI{2000}{\kelvin}. Thus, growth on rhenium NPs in these experiments likely proceeded via VSS rather than VLS. For NPs larger than $\sim\SI{2.5}{\nano\meter}$, a separation in melting behavior is observed, where icosahedral NPs exhibit significantly lower melting points than Wulff NPs, see Fig.~\ref{fig:cluster}c. This trend is consistent with Wulff NPs having lower free energy than icosahedral NPs above $\sim\SI{2.5}{\nano\meter}$ diameter, see Fig. S2. Polycrystalline NPs, on the other hand, are observed to anneal and evolve toward single-crystalline Wulff NPs, see SI Video 1, resulting in similar melting points.

Since the solubility of carbon in rhenium at typical CCVD temperatures is very low~\cite{arnoult1972solubility}, carbon atoms are expected to reside predominantly on the NP surface during growth rather than dissolve into the bulk. The effect of surface carbon on the melting point should therefore be considered. As shown in Fig.~\ref{fig:cluster}d, surface carbon does not significantly affect $T_\text{m}$ up to concentrations of $c = \SI{7.0}{\nano\meter^{-2}}$. During SWCNT growth, the surface carbon concentration is approximately $c = \SIrange{5.0}{6.0}{\nano\meter^{-2}}$ (shown later); therefore, the melting point of NPs during growth is not expected to be significantly affected.

A key difference between VLS and VSS growth is how carbon is transported to the growing tube. For VLS growth, carbon transport in the liquid catalyst occurs via diffusion and by bulk motion of the liquid (advection). However, for VSS growth, carbon transport is limited to bulk and/or surface diffusion and for rhenium, due to the low carbon solubility, transport is dominated by surface diffusion.

To quantify this, surface diffusion of carbon was evaluated on the facets that make up rhenium Wulff NPs. Two Wulff constructs were considered: a "clean" shape obtained for surface energies with no surface carbon, Fig.~\ref{fig:cluster}a, and a carbon-covered shape obtained for surface energies representative of growth conditions ($c = \SI{5.5}{\nano\meter^{-2}}$), Fig.~\ref{fig:cluster}b. As can be seen, the equilibrium shape of rhenium NPs changes significantly with the addition of surface carbon. However, since growth is out of equilibrium and is typically initiated from a comparatively clean NP, the instantaneous catalyst shape during growth is expected to lie between these. Nevertheless, the facets in Fig.~\ref{fig:cluster}a,b provide a consistent basis for comparing surface diffusion kinetics.

As shown in Fig.~\ref{fig:cluster}e, diffusion coefficients vary substantially across facets, with differences of approximately one to two orders of magnitude between the "fastest" facet $(0001)$ and the "slowest" facet $(10\bar{1}1)$, depending on temperature. This anisotropy is reflected in the facet-dependent activation energies shown in the inset of Fig.~\ref{fig:cluster}f. Relative to carbon diffusion in liquid iron~\cite{han1980determination}, a common VLS catalyst, surface diffusion on rhenium at \SI{1500}{\kelvin} is lower by a factor of $\sim 3$ on the $(0001)$ facet and by a factor of $\sim 50$ on the $(10\bar{1}1)$ facet, see Fig.~\ref{fig:cluster}f. Since the $(10\bar{1}1)$ facet dominates the carbon-covered rhenium NP during growth, see Fig.~\ref{fig:cluster}b, carbon delivery to the growing CNT is expected to be limited by diffusion on the $(10\bar{1}1)$ facet and therefore $\sim 50\times$ slower for VSS growth (on solid rhenium) compared to VLS growth (on liquid Fe).

\begin{figure*}
  \centering
  \includegraphics[scale=1.0]{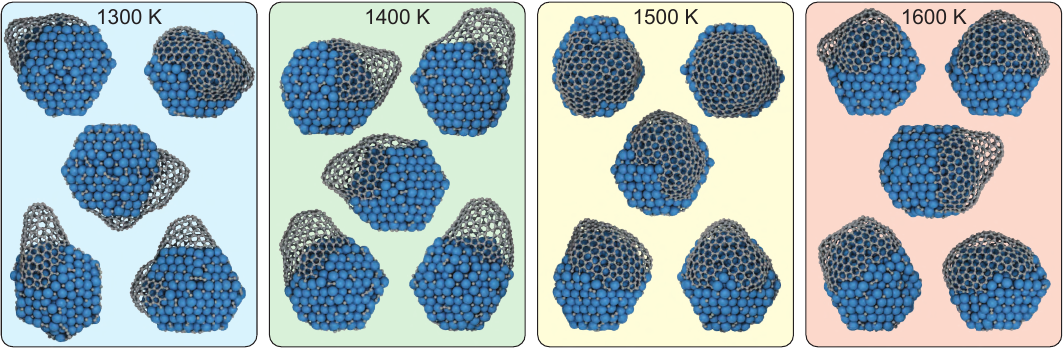}
  \caption{\label{fig:T_test} The impact of surface carbon diffusion, through changes in temperature, on cap nucleation and initial single-walled carbon nanotube (SWCNT) growth on rhenium nanoparticles (NPs). Growth was performed over \SI{6}{\micro\second} at a rate of $k = 50$ carbon atoms per \si{\micro\second} on a Re\textsubscript{594} NP. Here the structures obtained after \SI{6}{\micro\second} are shown where blue atoms depict rhenium and gray atoms carbon.}
\end{figure*}

The consequences of this sluggish carbon transport can, as a first approximation, be studied using diffusion on the surface of a sphere. For a $d = \SI{2.0}{\nano\meter}$ diameter particle, using a characteristic equator-to-pole diffusion distance $L = d\pi/4$ and a surface diffusion coefficient equal to that of the $(10\bar{1}1)$ facet at \SI{1500}{\kelvin}, $D_{(10\bar{1}1)}^{\text{1500 K}} = \SI{0.027}{\nano\meter^2\per\nano\second}$, gives a characteristic transport time of $L^2/(4D_{(10\bar{1}1)}^{\text{1500 K}}) = \SI{23}{\nano\second}$. Under idealized assumptions that incorporation into the CNT is instantaneous and that no carbon leaves the NP, this implies an upper-bound growth rate (carbon supply rate) of $k\approx44$ carbon atoms per \si{\micro\second}. Rates significantly exceeding this will result in carbon accumulation, which may give rise to further nucleation events or encapsulation.

For comparison, the growth rate of long, defect-free SWCNTs on liquid iron was found to be limited by interface defect formation and healing~\cite{hedman2024dynamics}. Previous work showed that at \SI{1500}{\kelvin}, growth rates up to 750 carbon atoms per \si{\micro\second} yield long defect-free tubes. This is an order of magnitude larger than the diffusion-limited estimate for rhenium (44 carbon atoms per \si{\micro\second}), highlighting that the kinetics of carbon transport, while negligible for VLS growth, play an important role in VSS growth.

\subsection{Nanotube nucleation and growth}

The sluggish carbon transport intrinsic to carbon-covered rhenium NPs is expected to affect VSS growth both in terms of cap nucleation and tube elongation. To investigate this, \si{\micro\second}-timescale growth simulations were performed on Re\textsubscript{594} Wulff NPs at a carbon supply (growth) rate of $k = 50$ carbon atoms per \si{\micro\second}, chosen to be close to the diffusion-limited estimate discussed above.

In practice, growth simulations with NEP and GPUMD on a Re\textsubscript{594} NP achieve up to \SI{150}{\nano\second} per day of simulated growth. At $k = 50$ carbon atoms per \si{\micro\second}, this corresponds to insertion of roughly seven carbon atoms per day. Consequently, complete growth simulations require multiple weeks even when initiated from NPs preloaded with surface carbon, underscoring that further improvements in the computational efficiency of MLIPs are needed.

The effect of surface diffusion of carbon, through changes in temperature, on cap nucleation and initial SWCNT growth was investigated using short \SI{6}{\micro\second} MD simulations (300 carbon atoms added) at \SIlist{1300;1400;1500;1600}{\kelvin} with five independent runs at each temperature. As shown in Fig.~\ref{fig:T_test}, growth at \SI{1300}{\kelvin} results in multiple nucleations or partial encapsulation. This behavior is consistent with the reduced surface diffusion coefficient at this temperature, $D_{(10\bar{1}1)}^{\text{1300 K}} = \SI{0.0055}{\nano\meter^2\per\nano\second}$. Using the same spherical diffusion model as previously yields a characteristic carbon transport time of $L^2/(4D_{(10\bar{1}1)}^{\text{1300 K}}) = \SI{111}{\nano\second}$, corresponding to approximately 9 carbon atoms per \si{\micro\second}. Therefore, at \SI{1300}{\kelvin} surface diffusion cannot sustain a growth (supply) rate of $k = 50$ carbon atoms per \si{\micro\second}. Increasing the temperature to \SI{1400}{\kelvin} results in single-cap nucleation and SWCNT growth. At this temperature, $D_{(10\bar{1}1)}^{\text{1400 K}} = \SI{0.013}{\nano\meter^2\per\nano\second}$, yielding a characteristic time of $L^2/(4D_{(10\bar{1}1)}^{\text{1400 K}}) = \SI{48}{\nano\second}$, or approximately 21 carbon atoms per \si{\micro\second}. This is within the same order of magnitude as the imposed growth rate and is sufficient to support growth over the simulated timescale.

At higher temperatures, where surface diffusion is faster, growth of SWCNTs is not observed. Instead, \SI{1500}{\kelvin} and above yield low-curvature graphitic structures consistent with encapsulation. Similar results were reported for VLS growth on iron at high temperatures and are attributed to the shorter lifetimes of interface defects~\cite{hedman2024dynamics}. In particular, reduced lifetimes of pentagon defects change the ratio of pentagons to hexagons, $N_5/N_6$, which lowers the Gaussian curvature $\propto N_5/N_6$, resulting in low-curvature graphitic structures. Therefore, although increasing temperature accelerates carbon transport, excessive temperature promotes encapsulation and suppresses CNT growth. Highlighting that, in contrast to VLS growth, the interplay between healing of interface defects and carbon transport is an important factor for VSS growth.


\begin{figure*}
  \centering
  \includegraphics[width=\textwidth]{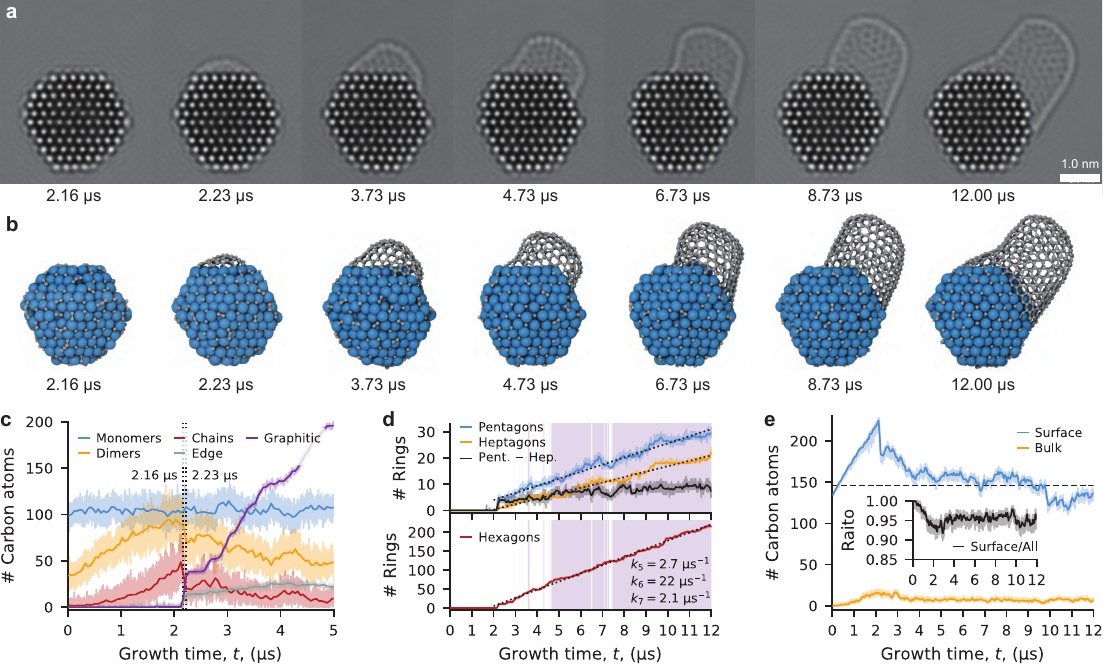}
  \caption{\label{fig:growth}\SI{12}{\micro\second} of simulated single-walled carbon nanotube growth on a Re\textsubscript{594} nanoparticle at \SI{1400}{\kelvin} and $k = 50$ carbon atoms per \si{\micro\second}. \textbf{a} simulated transmission electron microscopy images of the structures in \textbf{b} obtained from the growth trajectory. Here blue atoms depict rhenium and gray atoms carbon. \textbf{c} number of carbon atoms comprising each species during the early phases of growth. Transparent colored lines represent raw data and solid lines are time averages. Dotted vertical lines demarcate cap nucleation from \SIrange{2.16}{2.23}{\micro\second}. \textbf{d} evolution of pentagons and heptagons (top) and hexagons (bottom). The shaded purple region marks where $N_5 - N_7\geq 6$, where $N_5$ and $N_7$ are the numbers of pentagons and heptagons, respectively. $k_{5,6,7}$ denotes the average rate of formation of pentagons, hexagons, and heptagons, respectively, determined by the fitted dotted line. \textbf{e} distribution of "free" carbon atoms, i.e. carbon atoms that are not part of the tube. The inset shows the fraction of carbon on the surface.}
\end{figure*}

Based on the short \SI{6}{\micro\second} growth simulations presented above, \SI{1400}{\kelvin} and $k = 50$ carbon atoms per \si{\micro\second} yielded single-cap nucleation and SWCNT growth. Thus, to further study VSS growth, five independent \SI{12}{\micro\second}-long runs (600 carbon atoms added) were performed under the same conditions. Fig.~\ref{fig:growth}a,b show the results from one of these simulations that produced a well-defined, albeit defective, SWCNT.

Although the NP was initialized in the Wulff shape expected for a "clean" NP without surface carbon (Fig.~\ref{fig:cluster}a), it was preloaded with surface carbon and therefore reconstructs within the first \si{\micro\second}, see Fig. S3 and SI Video 2. By \SI{2.16}{\micro\second}, see the snapshot in Fig.~\ref{fig:growth}b, the NP acquires a shape consistent with the carbon-covered Wulff construct, with $(0001)$ and $(10\bar{1}1)$ facets dominating (Fig.~\ref{fig:cluster}b). This shows that, although the NP remains crystalline throughout growth, its exposed facets and overall shape evolve during SWCNT growth due to carbon adsorption.

Cap nucleation initiates at \SI{2.16}{\micro\second} on the $(0001)$ facet and proceeds rapidly, depleting the carbon on that facet within $\sim\SI{70}{\nano\second}$, second snapshot at \SI{2.23}{\micro\second} in Fig.~\ref{fig:growth}b. The species analysis in Fig.~\ref{fig:growth}c reflects this transition, showing a sharp decrease in carbon chains and dimers and a corresponding increase in graphitic and edge carbons between the vertical dashed lines, which resembles cap nucleation during VLS growth on iron~\cite{hedman2024dynamics,kohata2025edge} while differing by a slightly larger monomer population and substantially longer timescales, $\sim\SI{70}{\nano\second}$ for VSS versus $\sim\SI{5}{\nano\second}$ for VLS growth. Here, initial ring formation occurs through alignment of two long chains, followed by attachment of a dimer to an internal (backbone) site on one chain, forming a three-coordinated carbon atom. The resulting kink facilitates bonding to the neighboring chain, forming an initial hexagon that subsequently reconstructs into a pentagon with carbon chains attached, see Fig. S4 and SI Video 3.

As the cap grows, it reaches the edge of the $(0001)$ facet and anchors at the junction between facets. This is visible in the second snapshot at \SI{2.23}{\micro\second} in Fig.~\ref{fig:growth}b and in the simulated transmission electron microscopy (TEM) images, Fig.~\ref{fig:growth}a. Attachment at facet edges is favored because the facet-edge atoms are under-coordinated relative to atop atoms. Similar behavior has been reported experimentally during in situ TEM growth of SWCNTs on solid cobalt--tungsten, where the growing tube appeared attached to a facet edge~\cite{yang2022growth}.

For the five \SI{12}{\micro\second} growth simulations performed here, cap nucleation occurred on the $(0001)$ facet in two, on the $(10\bar{1}1)$ facet in one, and at a junction between facets in two runs. However, only two of the growth simulations yielded well-defined SWCNTs. In one, the cap formed on the $(0001)$ facet (Fig.~\ref{fig:growth}); in the other, it formed at the junction between the $(0001)$ and $(10\bar{1}1)$ facets. For the simulation discussed here, after nucleation the cap continued to grow and extended beyond the $(0001)$ facet, as seen in the snapshots \SIrange{3.73}{12.00}{\micro\second} in Fig.~\ref{fig:growth}b. A detailed look at the trajectory shows that the low curvature of the cap on the $(0001)$ facet enables dimers at the edge of the $(10\bar{1}1)$ facet to attach to the growing cap, forming rings that propagate the cap onto the $(10\bar{1}1)$ facet as seen in the snapshot at \SI{3.73}{\micro\second} in Fig.~\ref{fig:growth}b and SI Video 4. Thus, even if the cap nucleates atop one facet, growth is not necessarily limited to that facet if the cap has low curvature and extends to the facet edge.

\begin{figure*}
  \centering
  \includegraphics[width=\textwidth]{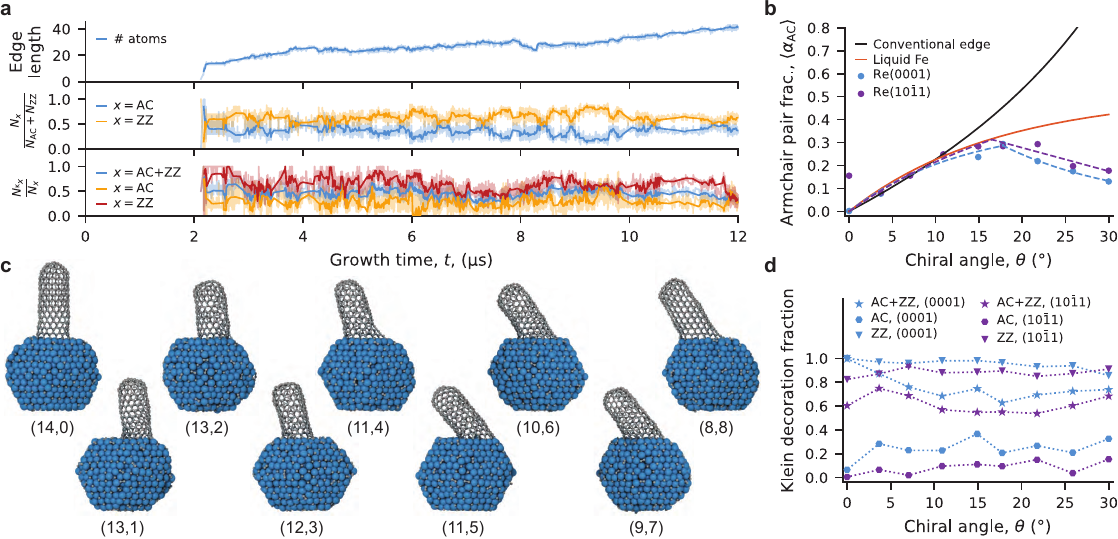}
  \caption{\label{fig:edge}The structure of the single-walled carbon nanotube (SWCNT) edge during growth and at equilibrium. \textbf{a} time series of the edge length (top), the fractions of armchair (AC) pairs, $N_\text{AC}$, and zigzag (ZZ) sites, $N_\text{ZZ}$ (middle), and the fractions of Klein-decorated AC (\textsuperscript{K}AC) and ZZ (\textsuperscript{K}ZZ) (bottom) for the growth simulation in Fig.~\ref{fig:growth}. \textbf{b} AC pair fraction, $\alpha_\text{AC}$, for SWCNTs on the $(0001)$ and $(10\bar{1}1)$ facets of a Re\textsubscript{1356} nanoparticle, averaged over the final \SI{500}{\nano\second} of a \SI{1}{\micro\second} annealing run at \SI{1600}{\kelvin}. \textbf{c} Final structures obtained after annealing for each chirality for tubes on the $(0001)$ facet. Here blue atoms depict rhenium and gray atoms carbon. \textbf{d} Fraction of Klein-decorated edge sites averaged over the final \SI{500}{\nano\second} of annealing.}
\end{figure*}

Although a significantly lower carbon supply rate was used here, $k = 50$ carbon atoms per \si{\micro\second}, compared to that used for defect-free VLS growth on iron, 500 carbon atoms per \si{\micro\second}~\cite{hedman2024dynamics}, defect-free growth was not observed. Instead, as shown in Fig.~\ref{fig:growth}d, the number of pentagons and heptagons in the tube increases continuously, indicating long-lived interface defects, i.e., slow defect healing. Here, pentagons and heptagons are incorporated into the tube at similar rates of $k_5 = 2.7$ and $k_7 = 2.1$ \SI{}{\micro\second^{-1}}, respectively. These rates are about one-tenth of the hexagon incorporation rate, $k_6 = \SI{22}{\micro\second^{-1}}$, and around five times larger than the defect incorporation rate reported for VLS growth on iron, $\sim\SI{0.5}{\micro\second^{-1}}$~\cite{hedman2024dynamics}. Given that the growth rate here is one-tenth that used for VLS growth on iron while the defect incorporation rate is five times higher, the interface defect lifetimes for VSS growth on rhenium are estimated to be about 50 times longer compared to VLS growth on iron.

For a three-coordinated network, the Euler/Gauss--Bonnet charge-balance rule requires $N_5 - N_7 = 6$ for a fully formed SWCNT-cap~\cite{dresselhaus1996science}, where $N_5$ and $N_7$ denote the numbers of pentagons and heptagons, respectively. Tracking the metric $N_5 - N_7\geq6$ over time, see the purple shaded region in Fig.~\ref{fig:growth}d, indicates that the cap becomes fully formed after \SI{7.48}{\micro\second}. This is substantially longer than the \SI{98.2}{\nano\second}, \SI{0.982}{\micro\second} after adjustment for the tenfold lower carbon supply rate, reported for VLS growth on iron and is consistent with less effective defect healing.

Finally, analysis of the free carbon shows that $\sim\qty{95}{\percent}$ resides on the NP surface during growth, see Fig.~\ref{fig:growth}e, consistent with the low carbon solubility in rhenium. Averaged over the last \SI{6}{\micro\second}, the carbon solubility in the Re\textsubscript{594} NP at \SI{1400}{\kelvin} is measured to be $1.13\pm0.24$ at\% in close agreement with the experimental value of $1.08$ at\% at \SI{1443.15}{\kelvin} for bulk rhenium~\cite{arnoult1972solubility}. Note that before cap nucleation, the measured solubility is higher due to the lack of a graphitic sink. Thus, the equilibrium carbon chemical potential is that of surface carbon rather than graphitic carbon.

\subsection{Nanotube interfaces at growth and equilibrium}

While growth simulations performed here resulted in defective tubes, the associated evolution of the SWCNT-edge can still provide insight into the interface kinetics of VSS growth.

As shown in Fig.~\ref{fig:edge}a (top), the edge length increases continuously during growth, consistent with incorporation of interface defects (Fig.~\ref{fig:growth}d). Nevertheless, by normalizing the number of armchair (AC) pairs, $N_\text{AC}$, and zigzag (ZZ) sites, $N_\text{ZZ}$, by the total number of edge species, $N_\text{AC} + N_\text{ZZ}$, it is clear that ZZ sites account, on average, for \qty{64}{\percent} of all edge sites, see Fig.~\ref{fig:edge}a (middle). This strong bias towards ZZ sites is consistent with the previously reported large difference between AC and ZZ interface energies for rhenium, \SI{0.33}{\electronvolt} per atom~\cite{hedman2019singlewalled}.

Additionally, both AC and ZZ edge atoms can be bonded to a third carbon atom, often called a Klein atom~\cite{klein1994graphitic}; here, this term denotes carbon atoms attached to either ZZ or AC atoms. The fractions of Klein-decorated AC and ZZ atoms, \textsuperscript{K}AC and \textsuperscript{K}ZZ, respectively, are shown in Fig.~\ref{fig:edge}a (bottom). On average, \qty{45}{\percent} of all edge atoms are Klein-decorated, but this strongly depends on the type, with only \qty{28}{\percent} of AC edge atoms being \textsuperscript{K}AC, while \qty{63}{\percent} of ZZ edge atoms are \textsuperscript{K}ZZ. This preference for ZZ sites and \textsuperscript{K}ZZ during VSS growth is also apparent in snapshots of Fig.~\ref{fig:growth}b and SI Video 5.

\begin{figure*}
  \centering
  \includegraphics[scale=1.0]{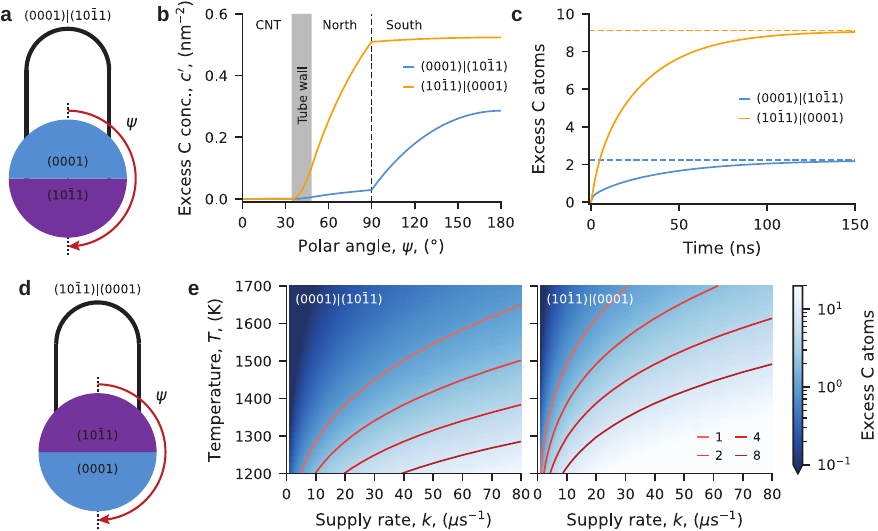}
  \caption{\label{fig:1d_diffusion} The effect of temperature and carbon supply rate on carbon concentration and excess carbon for carbon nanotube (CNT) growth on nanoparticles (NPs) with two facets having distinct diffusion coefficients. \textbf{a}, \textbf{b} simple model of a rhenium NP where the CNT grows on the $(0001)$ or $(10\bar{1}1)$ facet, respectively. Here the tube acts as a carbon sink and the carbon source is uniformly distributed on the surface of the NP outside the area covered by the tube. \textbf{c} steady-state carbon concentration, $c^\prime$; the dashed line marks the boundary between the facets. \textbf{d} excess carbon over time; the dashed lines denote the steady-state excess carbon. \textbf{e} map of the excess carbon atoms as a function of carbon supply rate (growth rate) and temperature. Here the contour lines show growth conditions resulting in 1, 2, 4 or 8 excess carbon atoms.}
\end{figure*}

To probe how solid rhenium NPs can affect the structure of the SWCNT-edge, tubes spanning chiral angles, $\theta = \tan^{-1}\!\left(\frac{\sqrt{3}m}{2n+m}\right)$, from \qtyrange{0}{30}{\degree} were attached to the $(0001)$ and $(10\bar{1}1)$ facets of Re\textsubscript{1356} NPs and annealed for \SI{1}{\micro\second} at \SI{1600}{\kelvin}. A higher annealing temperature compared to the growth temperature (\SI{1400}{\kelvin}) was used to reduce the time required to reach equilibrium.

Figure~\ref{fig:edge}b shows the AC pair fraction, $\langle\alpha_\text{AC}\rangle = N_\text{AC}/(N_\text{AC} + N_\text{ZZ})$, averaged over the last \SI{500}{\nano\second} of annealing for each chirality. At the beginning of each simulation, upon attachment to the NP, each SWCNT had a conventional edge corresponding to that of a perpendicularly cut tube. For such a conventional edge, $\alpha_\text{AC}(\theta) = \frac{2\tan\theta}{\sqrt{3}-\tan\theta}$ which is shown as the black line in Fig.~\ref{fig:edge}b. Although an initial etching stage is observed for all chiralities to achieve an equilibrium surface carbon concentration on the NP, see Fig. S5, tubes with small chiral angles, $\theta < \ang{15}$, still retain their conventional edge.

For SWCNTs on liquid catalysts with small to moderate differences in AC and ZZ interface energies, $\alpha_\text{AC}(\theta)$ follows the configurational entropy model of Bichara \emph{et al.}~\cite{magnin2018entropydriven,förster2023swinging,kohata2025edge}. This behavior is exemplified by equilibrium MD results for SWCNTs on liquid iron at \SI{1300}{\kelvin}~\cite{hedman2026chiral}, orange line in Fig.~\ref{fig:edge}b. However, for tubes on solid rhenium NPs, it is clear that $\alpha_\text{AC}(\theta)$ deviates from what is expected based on the configurational entropy, see Fig.~\ref{fig:edge}b. This suppression of AC pairs is most pronounced for $\theta\geq\ang{18}$ and is observed for tubes on both the $(0001)$ and $(10\bar{1}1)$ facets where $\langle\alpha_\text{AC}\rangle$ is significantly lower than for SWCNTs on liquid iron. As a result, strongly inclined and segregated interfaces are obtained for the $(11,5)$, $(10,6)$, $(9,7)$ and $(8,8)$ SWCNTs ($\theta\geq\ang{18}$), apparent in the annealed structures in Fig.~\ref{fig:edge}c. Here it is also clear that the tubes preferentially attach to facet edges.

These annealed structures also exhibit Klein-decorated edges, and averaging over the final \SI{500}{\nano\second} indicates that, on both facets, more than \qty{50}{\percent} of the edge atoms are Klein-decorated with only weak chirality dependence, see Fig.~\ref{fig:edge}d. Here a large fraction, more than \qty{80}{\percent} of the ZZ atoms, are \textsuperscript{K}ZZ, whereas fewer than \qty{40}{\percent} of the AC atoms are \textsuperscript{K}AC, again with only weak chirality dependence. A modest facet dependence is observed, however, where tubes on the $(0001)$ facet exhibit a stronger preference for both \textsuperscript{K}AC and \textsuperscript{K}ZZ.

\subsection{Diffusion-limited growth conditions}

MLIP-driven simulations show that VSS growth differs from VLS growth by (i) sluggish carbon transport regulated via facet-dependent surface diffusion and (ii) suppressed configurational entropy-driven edge structures. For rhenium, these effects manifest as slow growth and defect-healing kinetics together with stabilization of zigzag-rich, strongly Klein-decorated edges over a broad chirality range. To further study the effects of facet-dependent surface diffusion at temperatures and carbon supply (growth) rates inaccessible to the MLIP-driven simulations, a simplified diffusion model was employed.

Here the NP is approximated as a sphere with a diffusion coefficient, $D(\psi, T)$, that depends on polar angle, $\psi$, temperature, $T$, and is piecewise constant between two hemispheres. The growing CNT, which acts as a carbon sink, is located in the northern hemisphere (aligned with the $z$-axis), see Fig.~\ref{fig:1d_diffusion}a,b. No carbon atoms are supplied to the NP surface inside the tube, while the remaining exposed surface acts as a uniform carbon source so that the sink and source, $S(\psi)$, are balanced. Assuming invariance under azimuthal rotation, two-dimensional diffusion on the surface of a sphere reduces to one-dimensional diffusion on $0\leq\psi\leq\pi$ governed by
\begin{equation}
    \label{eq:1d_diffusion}
    \frac{1}{R^2\sin{\psi}}\frac{d}{d\psi}\left(D(\psi,T)\sin{\psi}\frac{dc^\prime}{d\psi}\right) + S(\psi) = 0,
\end{equation}
with the source term as
\[
    S(\psi)=
    \begin{cases}
        0, & 0\leq\psi<\psi_1\\
        -k/A_\text{sink}, & \psi_1\leq\psi\leq\psi_2\\
        +k/A_\text{source}, & \psi_2<\psi\leq\pi
    \end{cases}
\]
and boundary conditions
\[
    \frac{dc^\prime}{d\psi}\Big|_{\psi=0} = 0,\quad
    \frac{dc^\prime}{d\psi}\Big|_{\psi=\pi} = 0,\quad
    c^\prime(0) = 0.
\]
Here, the steady-state concentration, $c^\prime$, is determined up to an additive constant; thus $c^\prime$ can be interpreted as the excess surface carbon concentration relative to equilibrium.

At high carbon coverage, the $(0001)$ and $(10\bar{1}1)$ facets dominate the NP surface (Fig.~\ref{fig:cluster}b). These two facets have markedly different diffusion coefficients that, at \SI{1400}{\kelvin}, equal \SI{0.240}{\nano\meter^2\per\nano\second} for $(0001)$ and \SI{0.013}{\nano\meter^2\per\nano\second} for $(10\bar{1}1)$. Thus, two limiting cases are considered: growth on the $(0001)$ facet (fast diffusion) and growth on the $(10\bar{1}1)$ facet (slow diffusion), as shown in Fig.~\ref{fig:1d_diffusion}a,b. Using a sink rate equal to the carbon supply rate in the growth simulations, $k = 50$ carbon atoms per \si{\micro\second}, the model yields distinct steady-state carbon concentration profiles, see Fig.~\ref{fig:1d_diffusion}c. A higher excess concentration with a sharper gradient is obtained when growth occurs on the $(10\bar{1}1)$ facet compared to growth on the $(0001)$ facet. This difference is reflected in the time evolution of the excess carbon relative to equilibrium, Fig.~\ref{fig:1d_diffusion}d, which reaches a higher steady-state value for growth on the slow-diffusion facet.

It is worth noting here that in the absence of a sink (no growing CNT), a uniformly distributed source produces a uniform surface concentration. Therefore, the facet on which a CNT nucleates and grows strongly affects the distribution of carbon elsewhere on the NP. As growth progresses on a slow-diffusion facet, excess carbon accumulates on faster-diffusion facets, which can promote further nucleation events or encapsulation. This again shows a clear difference between VSS and VLS growth of CNTs.

The excess carbon on the NP also depends on the carbon supply rate, which is controlled experimentally by the feedstock decomposition (CCVD temperature and partial pressure). As seen in Fig.~\ref{fig:1d_diffusion}e, the excess carbon scales linearly with the carbon supply rate, as Eq.~\eqref{eq:1d_diffusion} is linear in $c^\prime$ and $S(\psi)\propto k$, but nonlinearly with the temperature due to the exponential dependence of the diffusivity (Fig.~\ref{fig:cluster}e). Thus, a very low carbon supply rate is needed in VSS growth to maintain low excess carbon and suppress multiple nucleation or encapsulation at low growth temperatures, especially when growth proceeds on a slow-diffusion facet. For experiments, this motivates a reduction in the carbon supply rate used, for example by lowering the feedstock partial pressure, to enable high-quality SWCNT growth under diffusion-limited VSS conditions.

\section{Conclusions}\label{sec:conclusions}

Machine-learning interatomic potential-driven molecular dynamics simulations on rhenium nanoparticles (NPs) indicate that single-walled carbon nanotube (SWCNT) growth proceeds in a vapor--solid--solid (VSS) regime over experimentally relevant temperature and NP sizes. Rhenium NPs remained crystalline with melting points well above typical growth temperatures, and surface carbon concentrations up to \SI{7.0}{\nano\meter^{-2}} do not substantially depress the melting point. Under these conditions, carbon transport is governed by surface diffusion rather than bulk dissolution, with diffusion coefficients varying by one to two orders of magnitude across NP facets. Carbon adsorption was further shown to drive dynamic changes in exposed facets and NP shape during growth while retaining crystallinity. As a consequence, sluggish, facet-dependent carbon transport imposes a diffusion-limited upper bound on sustainable carbon supply. Beyond this limit, excess surface carbon accumulates, resulting in encapsulation or additional nucleation events.

Consistent with diffusion-limited kinetics, \si{\micro\second}-timescale growth simulations revealed that single SWCNT-cap nucleation and growth occurred only within a narrow kinetic window, exemplified by \SI{12}{\micro\second} growth simulations at \SI{1400}{\kelvin} with 50 carbon atoms per \si{\micro\second}. Lower temperatures promoted carbon buildup leading to multiple nucleation and/or partial encapsulation, whereas higher temperatures favored low-curvature graphitic structures consistent with encapsulation, highlighting a competition between accelerated diffusion and shortened interface defect lifetimes. Cap nucleation occurred not only atop facets but also at facet junctions, where the cap/tube preferentially anchors at facet edges; subsequent growth proceeded at substantially lower rates than in vapor--liquid--solid (VLS) growth. Defect incorporation persisted during VSS growth, implying long-lived interface defects with lifetimes estimated to be $\sim\num{50}$ times longer than for VLS growth on iron, while equilibrium annealing indicated suppression of configurational entropy-driven edge structures and stabilization of zigzag-rich, strongly Klein-decorated edges across chiralities. Together with a simplified diffusion model, these results support a mechanistic picture in which the identity of exposed facets regulates carbon transport and growth kinetics, motivating stringent experimental control of carbon supply and temperature under diffusion-limited VSS conditions.

\section{Methods}\label{sec:methods}

\subsection{Neuroevolution potential}\label{subsec:nep}

The Neuroevolution potential (NEP) v4~\cite{song2024generalpurpose} was trained using GPUMD version 3.9.4. Training data were generated through active learning using an ensemble of NEPs. An ensemble of four NEPs was used to identify configurations encountered during molecular dynamics (MD) that were outside the distribution of the current training set by monitoring the deviation in the predicted forces (model deviation).

For a given configuration, the model deviation $\sigma_F$ was defined as the maximum force sample standard deviation over all atoms,
\[
\sigma_F = \max_i\sqrt{\sigma_{i,F_x}^2 + \sigma_{i,F_y}^2 + \sigma_{i,F_z}^2},
\]
where $\sigma_{i,F_k}^2$ (with $k\in\{x,y,z\}$) denotes the sample variance of the force component on atom $i$ computed over the four NEPs. Because $\sigma_F$ depends on the magnitude of the forces, it varies across systems and temperatures. A practical threshold $\tau_{\sigma_F}$ was selected in the range \SIrange{500}{1000}{\milli\electronvolt\per\angstrom}. Configurations with large model deviation ($\sigma_F>\tau_{\sigma_F}$) were extracted during MD, downselected using farthest-point sampling in the learned embedding space of local atomic environments, and labeled by density functional theory (DFT) calculations to obtain energies, forces, and virials. The newly labeled configurations were appended to the training set and a new NEP ensemble was trained. This procedure was repeated until the model deviation remained low ($\sigma_F<\tau_{\sigma_F}$) throughout the target MD simulations.

Active learning was performed for 18 iterations to reach an acceptable model deviation for test growth simulations. The final dataset comprised 96 023 images (Fig.~\ref{fig:sketch-map}) and was split into 90\% training and 10\% validation.

Local atomic environments were represented using 8 radial and 50 angular descriptors. Radial descriptors were constructed using 6 basis functions with a cutoff of \SI{5.8}{\angstrom}. Angular descriptors were constructed using 8 basis functions with a cutoff of \SI{4.2}{\angstrom}. The 50 angular descriptors were partitioned into 40 three-body and 10 four-body terms. Site energies were represented by a single-hidden-layer feedforward neural network acting on the descriptor vector. The hidden layer contained 58 neurons, equal to the input dimension ($8 + 50$).

Training was performed for 3 million generations (batches) using a population size of 64 and a batch size of 2000 structures. The loss function combined errors in energy $U$, forces $\mathbf{F}$, and virials $\mathbf{W}$,
\[
L = \lambda_e \frac{\|\hat{\mathbf{U}} - \mathbf{U}\|_2}{\sqrt{N}}
+ \lambda_f \frac{\|\hat{\mathbf{F}} - \mathbf{F}\|_2}{\sqrt{3M}}
+ \lambda_v \frac{\|\hat{\mathbf{W}} - \mathbf{W}\|_2}{\sqrt{6N}}.
\]
Here $\lambda_e$, $\lambda_f$, and $\lambda_v$ weight the energy, force, and virial contributions. For the first 1 million generations, uniform weights were used ($\lambda_e = 1.0$, $\lambda_f = 1.0$, $\lambda_v = 1.0$). The weights were then set to $\lambda_e = 1.0$, $\lambda_f = 0.1$, and $\lambda_v = 0.0$ for 1 million generations to emphasize energy accuracy. Because this weighting reduces force accuracy, a final 1 million-generation stage was performed with $\lambda_e = 1.0$, $\lambda_f = 1.0$, and $\lambda_v = 0.0$ to recover force accuracy while maintaining improved energy accuracy. For uncertainty estimation during MD, an ensemble of four NEP models was trained using the same procedure.

\subsection{Data labeling}\label{subsec:dft}

Each structure in the training dataset was labeled by single-point DFT calculations to obtain energies, forces, and virials. Calculations were performed using the Vienna \textit{Ab initio} Simulation Package (VASP)~\cite{kresse1993ab,kresse1996efficiency,kresse1996efficient} version 6.3.2. A plane-wave basis was employed together with the projector-augmented wave (PAW) method~\cite{blöchl1994projector,kresse1999ultrasoft} and standard pseudopotentials (\texttt{Re 17Jan2003} and \texttt{C 08Apr2002}).

Labeling was performed in two steps. First, loose DFT settings were used with the local density approximation (LDA) exchange-correlation functional and large initial magnetic moments of 5 $\mu_\text{B}$ for rhenium and 2 $\mu_\text{B}$ for carbon. Direct minimization using the preconditioned conjugate-gradient method (\texttt{ALGO = Conjugate}) was applied to obtain reliable magnetic moments. Second, these magnetic moments were used as the initial guess for calculations with tighter settings. The \texttt{rev-vdW-DF2} van der Waals exchange-correlation functional~\cite{klimeš2009chemical,hamada2014van} was used with non-spherical gradient corrections enabled (\texttt{LASPH = True}). A plane-wave cutoff energy of \SI{700}{\electronvolt} was applied (\texttt{ENCUT = 700}) and symmetry was disabled (\texttt{ISYM = 0}). The electronic self-consistent loop was converged to \SI{1e-6}{\electronvolt} (\texttt{EDIFF = 1.0E-6}). Gaussian smearing was used (\texttt{ISMEAR = 0}) with \SI{0.05}{\electronvolt} smearing width (\texttt{SIGMA = 0.05}). Spin polarization was enabled (\texttt{ISPIN = 2}) using the magnetic moments obtained from the first step. For periodic structures, a $\Gamma$-centered $k$-point mesh with density $\pi/10$ \SI{}{\angstrom^{-1}} was used (\texttt{KSPACING = 0.3142}). For non-periodic structures, only the $\Gamma$ point was used with at least \SI{15}{\angstrom} vacuum spacing between periodic images.

\subsection{Molecular dynamics}\label{subsec:md}

All MD simulations were performed using Graphics Processing Units Molecular Dynamics (GPUMD)~\cite{fan2017efficient,fan2022gpumd} version 3.9.4. One of the four trained NEPs (Sec.~\ref{subsec:nep}) was used to propagate each MD trajectory. The ensemble model deviation was evaluated every \SI{10.0}{\pico\second} as a measure of trajectory reliability. All simulations used a \SI{2.0}{\femto\second} timestep and a Nosé--Hoover chain thermostat with chain length 4.

\paragraph{Nanoparticle melting simulations.}
To determine the melting points in Fig.~\ref{fig:cluster}, nanoparticles were heated from \SI{1000}{\kelvin} to \SI{3500}{\kelvin} at \SI{25}{\kelvin\per\nano\second}. Wulff nanoparticles were constructed using DFT-calculated surface energies with and without a surface carbon concentration of $\SI{5.5}{\nano\meter^{-2}}$ for each of the 12 rhenium HCP facets available in the Crystalium database~\cite{tran2016surface}. Polycrystalline nanoparticles were obtained by first cooling from \SI{3500}{\kelvin} to \SI{1000}{\kelvin} at \SI{25}{\kelvin\per\nano\second} prior to heating. The melting point $T_\text{m}$ of each nanoparticle was identified from the discontinuity (first-order phase transition) in the potential energy during heating.

\paragraph{Surface diffusion simulations.}
Surface diffusion coefficients for carbon on rhenium HCP facets were computed using \SI{5}{\nano\meter} $\times$ \SI{5}{\nano\meter} supercell slabs of nine facets: $(0001)$, $(10\bar{1}0)$, $(11\bar{2}0)$, $(21\bar{3}0)$, $(10\bar{1}1)$, $(10\bar{1}2)$, $(21\bar{3}1)$, $(11\bar{2}1)$, and $(2\bar{1}\bar{1}2)$. Carbon atoms were randomly deposited on one side of the slab to reach a surface carbon concentration of $\SI{5.5}{\nano\meter^{-2}}$. Each system was equilibrated for \SI{10.0}{\nano\second} in the NPT ensemble at ambient pressure \SI{0.1}{\mega\pascal} using a Nosé--Hoover chain thermostat and a Parrinello--Rahman barostat. Production simulations were then performed for \SI{100}{\nano\second} in the NVT ensemble. Mean-squared displacements of carbon atoms were recorded every \SI{2.0}{\pico\second} and were used to extract diffusion coefficients.

\paragraph{Single-walled carbon nanotube growth simulations.}
SWCNT growth simulations were performed on Re\textsubscript{594} Wulff nanoparticles by inserting single carbon atoms every \SI{20}{\nano\second}, corresponding to an insertion (growth) rate of 50 carbon atoms per \si{\micro\second}. Carbon atoms were deposited at random surface locations selected to avoid unphysical insertion inside the nanotube or near the SWCNT/NP interface. A vector was first computed from the center of position of existing carbon atoms to the center of position of rhenium atoms. A conical sector was then defined with apex at the nanoparticle center and axis along this vector, pointing outward from the nanoparticle and away from the carbon cluster. The cone half-angle was \ang{90}. A random direction within this cone was sampled and, starting from the nanoparticle center, a point was marched outward along this direction until the distance to the nearest rhenium atom exceeded \SI{2.5}{\angstrom} and the distance to the nearest carbon atom exceeded \SI{1.4}{\angstrom}. A new carbon atom was then inserted at this position. To avoid excessive translation and rotation induced by repeated insertions, the total linear and angular momenta of the full system were zeroed every 50 timesteps.

\section{Acknowledgments}\label{sec:acknowledgments}

D. H. acknowledges financial support from the Institute for Basic Science (IBS-R019-D1). Many thanks to Dr. Bichara for insightful discussions and comments on the manuscript, merci !

\bibliography{main}

\end{document}